\documentclass[twocolumn,showpacs,preprintnumbers]{revtex4}
\usepackage{amsmath}
\usepackage{amssymb}
\usepackage{amsfonts}
\usepackage{graphicx}
\usepackage{dcolumn}
\usepackage{bm}

\input{tcilatex}

\begin{document}

\title{Quantum Games and the Relationships  \\ between Quantum Mechanics and
Game Theory}
\author{Esteban Guevara Hidalgo$^{\dag \ddag }$}
\affiliation{$^{\dag }$Center for Nonlinear and Complex Systems, Universit\`{a} degli
Studi \\
dell'Insubria, Via Valleggio 11, 22100 Como, Italy \\
$^{\ddag }$SI\'{O}N, Autopista General Rumi\~{n}ahui, Urbanizaci\'{o}n Ed\'{e}%
n del Valle,\\
Sector 5, Calle 1 y Calle A \# 79, Quito, Ecuador}

\begin{abstract}

Quantum games have proposed a new point of view for the solution of the
classical problems and dilemmas in game theory. It has been shown that are
more efficient than classical games and provide a saturated upper bound for
this efficiency. Certain quantization relationships can be proposed with the
objective that a game can be generalized into a quantum domain where the
linear superposition of actions is allowed. This quantization let us describe and solution problems originated by conflicting or cooperative behaviors among the members of a system from the point of
view of quantum mechanical interactions. This leads us to interesting aspects which only can be observed through the quantization of a game like the possibility of the entanglement between players, the definition of a socioeconomical temperature in a system and the analysis of a game through
elements of quantum information theory.

\smallskip

Although both systems analyzed are described through two theories apparently
different (quantum mechanics and game theory) both are analogous and thus
exactly equivalents. The quantum analogue of the replicator dynamics is the
von Neumann equation. The classical equilibrium concepts in game theory can
be also generalized through a maximum entropy approach in the so called
Collective Welfare Principle. Nature is a game in where its players compete
for the equilibrium of the system that they are
members. They act as a whole besides individuals like they obey a rule in
where they prefer to work for the collective besides the individual welfare.
If it is maximized the welfare of the individual above the collective
welfare the system gets unstable and eventually it collapses.

\smallskip

Quantum mechanics (and physics) could be used to explain more correctly
biological and economical processes (econophysics). A special consequence of the relationships between quantum mechanics and game theory is analyzed. It is shown that the so called \textquotedblleft globalization\textquotedblright\ process (i.e., the inexorable integration of markets, currencies, nation-states, technologies and the intensification of consciousness of the
world as a whole) has a behavior exactly equivalent to a system that is
tending to a maximum entropy state i.e., to its state of equilibrium. This let us predict the apparition of
big common markets and strong common currencies that will reach the
\textquotedblleft equilibrium\textquotedblright\  by decreasing its number until they get a state characterized
by only one common currency and only one big common community around the
world.

\end{abstract}

\maketitle
\email{esteban\_guevarah@yahoo.es}

\section{Introduction} 

The present work offers an overview of game theory. Starting from the classical theory the basic concepts of game, strategy, equilibrium, the evolutionary theory, the replicator dynamics and evolutionary stable strategies and a review of the most important works in the recent field of quantum game theory, its importance, and how the quantization can improve the results in a game and solution the dilemmas of game theory. Finally, we analyze the relationships between game theory and quantum mechanics, its consequences and applications.

\newpage	 

\section{Game Theory: From Nash to the Replicator Dynamics}

\subsection{Classical Game Theory}

Game theory \cite{1,2,3} is the study of decision making of competing agents
in some conflict situation. It tries to understand the birth and development of conflicting or cooperative behaviors among a group of individuals who behave rationally and strategically according to their
personal interests. Each member in the group strive to maximize its welfare,
state, utilities or payoffs by choosing the best courses of strategies from
a cooperative or individual point of view. Game theory has been applied to
solve many problems in economics, social sciences, biology, computer
science, international relations, engineering and more recently, in physics.

\subsubsection{Basic Definitions}

A \emph{game} $G=(N,S,E)$ consists of a set of players $N$, a set of
strategies $S=\left\{ S_{1},...,S_{N}\right\} $, where $S_{j}$ is the set of
strategies available to the $j$th player and a set of payoff functions 
$E=\left\{ E_{1},...,E_{N}\right\}$, where $E_{j}$ is the payoff function for the $j$th player. A \emph{payoff function} $E$ for a player is a mapping from the
cross-product of player's strategy spaces to the player's set of payoffs. $E$
assigns a real number to the pair $(s_{i},s_{j})$, $E(s_{i},s_{j})$ is the
payoff obtained by a player who plays the strategy $s_{i}$\ against an
opponent who plays the strategy $s_{j}$. An \emph{action} or a move is a choice available to a player. It could be taken by a player during some moment in a game.

\subsubsection{Strategies}

A \emph{strategy} is a complete plan of action for every stage of the game,
regardless of whether that stage actually arises in play. A \emph{strategy
space} for a player is the set of all strategies available to the player. A \emph{pure strategy} is a strategy that specifies a unique move in a given game position, i.e., an action
with probability 1. A \emph{mixed strategy} $x$ is a probability distribution over $S$ which corresponds to how frequently each move is chosen. A \emph{dominant strategy} is a strategy that does at least as well as any competing strategy against any possible moves by the other player(s).

\subsubsection{Games}

A \emph{zero sum game} is a game in where the sum of all players payoffs is zero regardless of the
strategies they choose. A player gains only at the expense of others. In a \emph{constant-sum game}, the sum of all players' payoffs is the same for any
outcome. A \emph{cooperative game} is a game in which two or more players strive toward a unique objective and therefore win or lose as a group. In a \emph{non-cooperative game} no outside authority assures that players stick to the same predetermined rules, and so binding agreements are not feasible. In these games players may cooperate but any cooperation must be self-enforcing. In a \emph{game of perfect information} the knowledge about other players is available to all participants i.e., every player knows the payoff functions and the strategies available to other players. A \emph{symmetric game} is a game in where all agents have the same set of strategies and identical payoffs functions, except for the interchange of roles of the players. A \emph{symmetric two-person game}  $G=(S,E)$ consists of a
finite nonempty pure strategy set $S$ and a payoff function $E$ which assigns a real number to the pair $(s_{i},s_{j})$. A $n_{1}\times n_{2}\times ...\times n_{N}$ game is a $N$ player game where the $j$th player has available $n_{j}$ strategies.

\subsubsection{Equilibrium Notions}

Let be $p,r\in S_{i}$ and $q\in S_{j}$. A \emph{best reply} to $q$
is a strategy $p$ which maximizes $E(p,q)$. A \emph{dominant strategy
equilibrium} is a strategy profile in which each player plays best replies
that do not depend on the strategies of other players. An \emph{equilibrium point} is a pair $(p,q)$ with the property that $p$ and $q$ are
best replies to each other. A strategy $r$ is a \emph{strict best reply} to
a strategy $q$ if it is the only best reply to $q$. A strict best reply must
be a pure strategy. An equilibrium point $(p,q)$ is called \emph{strict
equilibrium point} if $p$ and $q$ are strict best replies to each other. A
best reply to $p$ which is different from $p$ is called \emph{alternative
best reply}.

A \emph{Nash equilibrium} (NE) \cite{4,5} is a set of strategies,
one for each player, such that no player has an incentive to unilaterally
change his action. Players are in equilibrium if a change in strategies by
any one of them would lead that player to earn less than if he remained with
his current strategy. A Nash equilibrium satisfies the following condition%
\begin{equation}
E(p,p)\geq E(r,p)\text{.} \label{1}
\end{equation}%
A player cannot increase his payoff if he decides to play the strategy $r$
instead of $p$. A \emph{focal point} is one amongst several NE which for
psychological reasons is particularly compelling.

In a zero-sum game between players A and B, player A should attempt to
minimize player B's maximum payoff while player B attempts to maximize his
own minimum payoff. When they do so the minimum of the maximum (minimax)
payoffs equals the maximum of the minimum (maximin) payoffs. Neither player
can improve his position, and so these strategies form an equilibrium of the
game. The \emph{minimax theorem} \cite{1} states that for every two-person,
zero-sum game, there always exists a mixed strategy for each player such
that the expected payoff for one player is the same as the expected cost for
the other. In other words, there is always a rational solution to a
precisely defined conflict between two people whose interests are completely
opposite. It is a rational solution in that both parties can convince
themselves that they cannot expect to do any better, given the nature of the
conflict.

A \emph{Pareto optimal} (PO) \cite{6} is a game result from which no player
can improve their payoff without another player being worse off, that is, if 
$\forall k,$ $\exists l$ such that%
\begin{gather}
E_{k}(s_{1},...,s_{k}^{\prime
},s_{l},...,s_{N})>E_{k}(s_{1},...,s_{k},s_{l},...,s_{N})\text{,}  \notag \\
\text{then }E_{l}(s_{1},...,s_{k}^{\prime
},s_{l},...,s_{N})<E_{l}(s_{1},...,s_{k},s_{l},...,s_{N}) \text{.} \label{2}
\end{gather}%
Then the unprimed strategy profile is Pareto optimal. An outcome of a game
is Pareto optimal if there is no other outcome that makes every player at
least as well off and at least one player strictly better off. That is, a
Pareto optimal outcome cannot be improved upon without hurting at least one
player.

\subsection{Evolutionary Game Theory}

Evolutionary game dynamics is the application of population dynamical
methods to game theory. It has been introduced by evolutionary biologists,
anticipated in part by classical game theorists \cite{7} and first
introduced under the name of \emph{evolutionary game theory} by J. Smith and G. Price in the context of animal conflict \cite{8}. Evolutionary game theory \cite{7,9,10} does not rely on rational assumptions (like
classical game theory) but on the idea that the Darwinian process of natural
selection \cite{11} drives organisms towards the optimization of
reproductive success \cite{12}. It combines the principles of game theory,
evolution, non linear dynamics and dynamical systems to explain the
distribution of different phenotypes in biological populations. Instead of
working out the optimal strategy, the different phenotypes in a population
are associated with the basic strategies that are shaped by trial and error
by a process of natural selection or learning. \emph{%
Strategies} are considered to be inherited programs for any conceivable
situation which control the individual's behavior. The members of a
population interact in game situations and the joint action of mutation and
selection replaces strategies by others with a higher reproductive success.
In this kind of games is less important to know which member plays which
strategy within a population but it is important to know the relative
frequency of actions (the probability of playing a strategy) \cite{12}. \emph{Payoffs} in biological
games are in terms of fitness a measure of reproductive success.

In contrast with classical game theory its evolutionary version deals with
entire populations of players, all \textquotedblleft
programmed\textquotedblright\ to use some strategy (or type of behavior).
Strategies with high payoff will spread within the population (this can be
achieved by learning, by copying or inheriting strategies, or even by
infection). The payoffs depend on the actions of the coplayers and hence on
the frequencies of the strategies within the population. Since these
frequencies change according to the payoffs, this yields a \emph{feedback
loop }\cite{7}. The dynamics of this feedback loop is the object of
evolutionary game theory. The feedback dynamics depend strongly, of course,
on the population structure, on the underlying game and on the way
strategies spread. Thus there are many \textquotedblleft game
dynamics\textquotedblright, which can be discrete or continuous, stochastic
or deterministic \cite{7,10,13,14}.

Many successful applications of evolutionary game theory appeared in
mathematical biology to explain biological phenomena (e.g., to predict the
behavior of bacteria and insects) but it can also be used to interpret
classical games from a different perspective. Instead of directly
calculating properties of a game, populations of players using different
strategies are simulated and a process similar to natural selection is used
to determine how the population evolves. This is made through the stability
analysis of differential equations and the implications to the games \cite{13}.

\subsubsection{Evolutionary Stable Strategies}

The central equilibrium concept of evolutionary game theory is the notion of
Evolutionary Stable Strategy (ESS) introduced by J. Smith and G. Price \cite%
{8,9}. An ESS is described as a strategy which has the property that if all
the members of a population adopt it, no mutant strategy could invade the
population under the influence of natural selection. ESS are interpreted as
stable results of processes of natural selection. The natural selection
process that determines how populations playing specific strategies evolve
is known as the replicator dynamics \cite{14,7,10,13} whose stable fixed
points are Nash equilibria \cite{2}.

Each agent in a n-player game (where the $i$th player has as strategy space $%
S_{i}$) is modeled by a population of players which have to be partitioned
into groups. Individuals in the same group would all play the same strategy. Randomly, we make play the members of the subpopulations against each other.
The subpopulations that perform the best will grow and those that do not
will shrink and eventually will vanish. The process of natural selection
assures survival of the best players at the expense of the others. A
population equilibrium occurs when the population shares are such that the
expected payoffs for all strategies are equal.

Consider a large population in which a two person game $G=(S,E)$ is played
by randomly matched pairs of animals generation after generation. Let $p$ be
the strategy played by the vast majority of the population, and let $r$ be
the strategy of a mutant present in small frequency. Both $p$ and $r$ can be
pure or mixed. An \emph{evolutionary stable strategy} (ESS) $p$ of a
symmetric two-person game $G=(S,E)$ is a pure or mixed strategy for $G$
which satisfies the following two conditions%
\begin{gather}
E(p,p)>E(r,p)\text{,}  \notag \\
\text{If }E(p,p)=E(r,p)\text{ then }E(p,r)>E(r,r)\text{.}  \label{3}
\end{gather}%
Since the stability condition only concerns to alternative best replies, $p$
is always evolutionarily stable if $(p,p)$ is an strict equilibrium point.
An ESS is also a Nash equilibrium since it is the best reply to itself and
the game is symmetric. The set of all the strategies that are ESS is a
subset of the NE of the game. A population which plays an ESS can withstand
an invasion by a small group of mutants playing a different strategy. It
means that if a few individuals which play a different strategy are
introduced into a population in an ESS, the selection process would
eventually eliminate the invaders.

\subsubsection{The Replicator Dynamics}

The natural selection process that determines how populations playing
specific strategies evolve is known as the \emph{replicator dynamics}. 
It describes the evolution of a polymorphic state in a population
represented by a mixed strategy $x$ for $G$ whose members are involved in a
conflict described by a symmetric two-person game $G=(S,E)$. The
probability assigned to a pure strategy $s$ is denoted by $x(s)$. If $%
s_{i},i=1,...,n\in S$ are the pure strategies available to a player, then
the \emph{player's strategy} will be denoted by the column vector $x$ with $%
x_{i}\in \left[ 0,1\right]$ and $\sum_{i=1}^{n}x_{i}=1$. The $i$th
component of $x$ gives the probability of playing strategy $s_{i}$ and also
is interpreted as the relative frequency of individuals using strategy $%
s_{i}$. Playing a pure strategy $s_{j}$ is represented by the vector $x$
whose $j$th component is $1$, and all the other components are 0. The 
\emph{fitness function} $E=f_{i}(x),$ $i=1,...,n$ specifies how successful
each subpopulation is and must be defined for each component of $x$. The
fitness for $x_{i}$ is the expected utility of playing strategy $s_{i}$
against a player with a mixed strategy defined by the vector $x$. It is
given by%
\begin{equation}
f_{i}(x)=(Ax)_{i}=\sum_{j=1}^{n}a_{ij}x_{j}\text{,}  \label{4}
\end{equation}%
where $A$ is the payoff matrix ($a_{ij}$ are its elements) and the subscript 
$i=1,...,n$ in $(Ax)_{i}$ denotes the $i$th component of the matrix-vector
product $(Ax)$. The \emph{average fitness of the population} $\left\langle
f(x)\right\rangle =\sum_{i=1}^{n}x_{i}f_{i}(x)$ is%
\begin{equation}
\left\langle f(x)\right\rangle =x^{T}Ax=\sum_{k,l=1}^{n}a_{kl}x_{k}x_{l}%
\text{,}  \label{5}
\end{equation}%
where the superscript $T$ denotes transpose.

The evolution of relative frequencies in a population is described by the 
replicator dynamics%
\begin{gather}
\frac{dx_{i}(t)}{dt}=\left[ f_{i}(x)-\left\langle f(x)\right\rangle \right]
x_{i}(t) \text{,} \notag
\\
\frac{dx_{i}(t)}{dt}=\left[ (Ax)_{i}-x^{T}Ax\right] x_{i}(t)\text{,}  \label{6}
\end{gather}%
or also%
\begin{equation}
\frac{dx_{i}(t)}{dt}=\left[ \sum_{j=1}^{n}a_{ij}x_{j}-%
\sum_{k,l=1}^{n}a_{kl}x_{k}x_{l}\right] x_{i}(t)\text{.} \label{7}
\end{equation}%
The stable fixed points of the replicator dynamics are Nash equilibria. It
is important to note that the fixed points of a system do not change in the
time. It means that if a population reaches a state which is a Nash
equilibrium, it will remain there. The replicator dynamics rewards
strategies that outperform the average by increasing their frequency, and
penalizes poorly performing strategies by decreasing their frequency.

In a symmetric game payoff matrices and actions are identical for both
agents. These games can be modeled by a single population of individuals
playing against each other. When the game is \emph{asymmetric}, a different
population of players must be used to simulate each agent. The strategy
vector for player one is represented by $x$ and for player two is
represented by $y$. Player one has $n$ strategies $s_{1i}\in S_{1},i=1,...,n$ and player two has $m$ strategies $s_{2j}\in S_{2},j=1,...,m$. Each player will have also a distinct payoff matrix $A\in 
\mathbb{R}
^{n\times m}$ and $B\in 
\mathbb{R}
^{m\times n}$, respectively. The fitness for a player who plays the
strategy $s_{1i}$ will be $f_{1i}=(Ay)_{i}$ and the average fitness of
the first population will be $\left\langle f_{1}\right\rangle =x^{T}Ay$.
Similarly, the fitness for a player who plays the strategy $s_{2i}$ will
be $f_{2i}=(Bx)_{i}$ and the average fitness of the second population will
be $\left\langle f_{2}\right\rangle =y^{T}Bx$. The evolution of this game
would be described for the next equations system \cite{7}%
\begin{gather}
\frac{dx_{i}(t)}{dt}=\left[ (Ay)_{1}-x^{T}Ay\right] x_{i}(t)\text{,}  \notag
\\
\frac{dy_{i}(t)}{dt}=\left[ (Bx)_{1}-y^{T}Bx\right] y_{i}(t)\text{.}
\label{8}
\end{gather}

\section{Quantum Mechanics Foundations}

\subsection{Physical System Representation}

Lets represent a physical system through a Hilbert space. A state of that system is completely described through a state vector $\left\vert \Psi(t)\right\rangle$ (element of that Hilbert space) and it is postulated that contains all the information about that system. The possible configurations of a physical system are described by the state space which is spanned by a set of basis states $B$. The number of basis states equals the dimension of the state space which can be infinite and even uncountable. A basis state $i \in B$ is denoted by $\left\vert i\right\rangle$ and every state $\left\vert \Psi\right\rangle $ can be described with a linear combination on their basis states $B$ 
\begin{equation}
\left\vert \Psi(t)\right\rangle= \sum_{i \in B}\alpha_{i}\left\vert i\right\rangle
\label{9}
\end{equation}
with $\alpha_{i} \in
\mathbb{C}
$. 
It can be shown that the inner product in the Hilbert space relates to the probability amplitudes in quantum mechanics and viceversa. By definition%
\begin{equation}
\left\vert \Psi(t)\right\rangle = \sum_{i \in B}\left\vert i\right\rangle \left\langle i\left\vert \Psi \right. \right\rangle \text{.} \label{10}
\end{equation}%
The probability of measuring a state $\left\vert \Psi\right\rangle$ in the basis state $i$ equals $\vert \left\langle i\left\vert \Psi \right. \right\rangle\vert^{2}$ which satisfies a normalization condition $\sum_{i\in B}\vert \left\langle i\left\vert \Psi \right. \right\rangle\vert^{2}=1$ for every $\left\vert \Psi\right\rangle$ in the Hilbert space.

In classical mechanics we can precisely specify the state of a system by one point in its phase space. Its trajectory through the phase space describes the time evolution of the system and this evolution follows Newton's laws or Hamilton equations. When the information about the system is incomplete the state of a system is not perfectly
defined for which we have to describe our system in terms of probabilities. 

An \emph{ensemble} is a collection of identically prepared physical
systems. When each member of the ensemble is characterized by the same state
vector $\left\vert \Psi (t)\right\rangle$ it is called \emph{pure ensemble}. If each member has a probability $p_{i}$ of being in the state $%
\left\vert \Psi_{i}(t)\right\rangle$ we have a \emph{mixed ensemble}. Each member of a \emph{mixed ensemble} is a pure state and its evolution is given by Schr\"{o}dinger equation. Probabilities for each state are constrained to satisfy the normalization condition $\sum_{i=1}^{n}p_{i}=1$  and obviously $0\leq p_{1},p_{2 },...,p_{n}\leq 1.$

An \emph{observable} is a property of the system state that can be determined by some sequence of physical operations.  Observables in our model are represented through  Hermitian (or self adjoint) operators acting over the Hilbert space. An operator $A$ is Hermitian when  $A=A^\dag$. Every operator can be represented in matrix form in some basis. In quantum physics we can know only the expectation value of an observable. Suppose we make a \emph{measurement} on a mixed ensemble of some observable $A$. The \emph{ensemble average} of $A$ is defined by the average of the expected values measured in each member of the ensemble described by $\left\vert \Psi _{i}(t)\right\rangle$ and with probability $p_{i}$, $\left\langle A\right\rangle _{\rho }=p_{1}\left\langle 
A\right\rangle _{1}+p_{2}\left\langle A\right\rangle
_{2}+...+p_{n}\left\langle A\right\rangle _{n}$%
\begin{gather}
\left\langle A\right\rangle_{\rho }=\sum_{i=1}^{n}p_{i}\left\langle \Psi_{i}(t)\right\vert A\left\vert \Psi _{i}(t)\right\rangle \text{,} \notag \\
\left\langle A\right\rangle_{\rho }=\sum_{i,j,k=1}^{n}p_{i}a_{jk}c_{j}^{(i)\ast }(t)c_{k}^{(i)}(t)\text{,} \label{11}
\end{gather}%
where $a_{jk}$ are the elements of the matrix that represents the observable 
$A$. The terms $c_{k}^{(i)}(t)=\left\langle k\left\vert \Psi _{i}(t)\right.
\right\rangle$ and $c_{j}^{(i)\ast }(t)=\left\langle \Psi _{i}(t)\left\vert
j\right. \right\rangle $ are the elements of certain \emph{density operator} $\rho (t)$ defined as
\begin{equation}
\rho (t)=\sum_{i=1}^{n}p_{i}\left\vert \Psi _{i}(t)\right\rangle
\left\langle \Psi _{i}(t)\right\vert\text{.}\label{12}
\end{equation}%
For a mixed state $\rho$ is $Hermitian$, $Tr\rho (t)=1$, $\rho ^{2}(t)\leq \rho (t)$ and $Tr\rho ^{2}(t)\leq 1$. 

To describe correctly a statistical mixture of states it is necessary the introduction of the density operator.  It contains all the physically significant information we can obtain about the ensemble in question. Any two ensembles that produce the same density operator are physically indistinguishable. A pure state is specified by $p_{i}=1$ for some $\left\vert \Psi_{i}(t)\right\rangle$, $i=1,...,n$ and its density operator $\rho (t)$ by a matrix with all its elements equal to zero except one 1 on the diagonal.
The diagonal elements $\rho _{nn}$ of the density operator $\rho (t)$
represents the average probability of finding the system in the state $%
\left\vert n\right\rangle $.%
\begin{gather}
\rho _{nn} =\left\langle n\right\vert \rho (t)\left\vert n\right\rangle 
=\sum_{i=1}^{n}\left\langle n\left\vert \Psi _{i}(t)\right. \right\rangle
p_{i}\left\langle \Psi _{i}(t)\left\vert n\right. \right\rangle \text{,} \notag \\
\rho _{nn}=\sum_{i=1}^{n}p_{i}\left\vert c_{n}^{(i)}\right\vert ^{2} \text{,} \label{13}
\end{gather}%
where $c_{n}^{(i)}=\left\langle n\left\vert \Psi i(t)\right. \right\rangle $
and $\left\vert c_{n}^{(i)}\right\vert ^{2}\in 
\mathbb{R}
^{+}$. If the state of the system is $\left\vert \Psi _{i}(t)\right\rangle $%
, $\left\vert c_{n}^{(i)}\right\vert ^{2}$ is the probability of finding, in
a measurement, this system in the state $\left\vert n\right\rangle $. The
diagonal elements $\rho _{nn}$ are zero if and only if all $\left\vert
c_{n}^{(i)}\right\vert ^{2}$ are zero. The non-diagonal elements $\rho _{np}$
expresses the interference effects between the states $\left\vert
n\right\rangle $ and $\left\vert p\right\rangle $ which can appear when the
state $\left\vert \Psi _{i}\right\rangle $ is a coherent linear
superposition of these states.%
\begin{gather}
\rho _{np} =\left\langle n\right\vert \rho (t)\left\vert p\right\rangle 
=\sum_{i=1}^{n}\left\langle n\left\vert \Psi _{i}(t)\right. \right\rangle
p_{i}\left\langle \Psi _{i}(t)\left\vert p\right. \right\rangle \text{,} \notag \\
\rho _{np} =\sum_{i=1}^{n}p_{i} c_{n}^{(i)}(t)c_{p}^{(i)\ast }(t) \label{14}
\end{gather}%
with $c_{n}^{(i)}(t)=\left\langle n\left\vert \Psi _{i}(t)\right.
\right\rangle $, $c_{p}^{(i)\ast }(t)=$ $\left\langle \Psi _{i}(t)\left\vert
p\right. \right\rangle $ and $c_{n}^{(i)}c_{p}^{(i)\ast }\in 
\mathbb{C}
$. If $\rho _{np}=0$ it means that the average has cancelled out any
interference effects between $\left\vert n\right\rangle $ and $\left\vert
p\right\rangle $ but if it is different from zero subsists certain coherence
between these states.

\subsection{Evolution of a Physical System}
\subsubsection{Schr\"{o}dinger \& von Neumann Equations}
Each pure state evolves following the Schr\"{o}dinger equation but the evolution of the system as a statistical mixture of states described through a density operator is given by the von Neumann
equation%
\begin{equation}
i\hbar \frac{d\rho}{dt}=\left[\hat{H},\rho \right] \text{,} \label{15}
\end{equation}%
$\hat{H}$ is the Hamiltonian of the physical system. The von Neumann
equation is only a generalization (and/or a matrix/operator representation) of the
Schr\"{o}dinger equation and the quantum analogue of Liouville's theorem
from classical statistical mechanics.

\subsubsection{Unitary Operators}
The evolution of an isolated quantum system is also described by unitary transformations ($U$ is unitary if $U^{\dag} = U^{-1}$). The  states $\left\vert \Psi_{1}\right\rangle$ at time $t_{1}$ and $\left\vert \Psi_{2}\right\rangle$ at time $t_{2}$ are related by an unitary transformation $U$ by%
\begin{equation}
\left\vert \Psi_{2}\right\rangle = U \left\vert \Psi_{1}\right\rangle \text{.} \label{16}
\end{equation}%
For a statistical mixture of states the ensemble evolves unitarily in time by%
\begin{equation}
\rho(t_{2})=U(t_{2},t_{1})\rho(t_{1})U^{\dag}(t_{2},t_{1}) \text{.}  \label{17}
\end{equation}%

\subsection{Quantum Bits and Quantum Registers}
A general \emph{qubit} state in a two-dimensional Hilbert space whose orthonormal basis can be written as 
$\left\{  \left\vert 0\right\rangle, \left\vert 1\right\rangle \right\}$ is%
\begin{equation}
\left\vert \Psi \right\rangle=a\left\vert 0\right\rangle + b\left\vert 1\right\rangle \label{18}
\end{equation}%
with $a$, $b$ $\in 
\mathbb{C}
$ and  $\vert a\vert^{2} + \vert b\vert^{2}=1$. In other words, $\left\vert \Psi\right\rangle$ is a unit vector in a two-dimensional complex vector space for which a particular basis has been fixed. If we expand this state space to that of a system whose basis set is described by $\left\{0, 1 \right\}^{n}$  we get the definition of a n-qubit system. The possible configurations of such a \emph{quantum register} are covered by%
\begin{equation}
\left\vert \Psi\right\rangle = \sum_{i \in \left\{0,1\right\}^{n}} \alpha_{i}\left\vert i \right\rangle\text{.}\label{19}
\end{equation}%
The state space of a n-qubit system equals the tensor product of $n$ separate qubit systems $\mathcal{H}_{\left\{0,1\right\}^{n}}=\mathcal{H}_{\left\{0,1\right\}}\otimes \mathcal{H}_{\left\{0,1\right\}} \otimes...\otimes\mathcal{H}_{\left\{0,1\right\}}$.
Qubits and quantum registers are used to describe the memory of quantum computers. 

\subsection{Quantum Entanglement}

Consider a system which has associated a Hilbert space $\mathcal{H}$ that can be divided into two subsystems. Assume $\mathcal{H}_{A}$ and $\mathcal{H}_{B}$ to be the Hilbert spaces corresponding to the subsystems $A$ and $B$, respectively. The two state spaces $A$ and $B$ are spanned by $\left\vert i\right\rangle_{A}$ and $\left\vert  j\right\rangle_{B}$. The state space is expanded into $\mathcal{H}$ through the tensor product $\otimes$ of $\mathcal{H}_{A}$ and $\mathcal{H}_{B}$ i.e., $\mathcal{H}=\mathcal{H}_{A}\otimes\mathcal{H}_{B}$. This space is spanned by the basis vectors $\left\vert k\right\rangle=\left\vert i\right\rangle_{A}\otimes \left\vert j\right\rangle_{B}$ sometimes denoted $\left\vert  i\right\rangle_{A}\left\vert  j\right\rangle_{B}$,  $\left\vert i,j\right\rangle_{AB}$ or $\left\vert ij\right\rangle_{AB}$. Any state $\left\vert \Psi\right\rangle_{AB}$ of $\mathcal{H}$ is a linear combination of the basis states $\left\vert i\right\rangle_{A}\left\vert j\right\rangle_{B}$%
\begin{gather}
\left\vert \Psi\right\rangle_{AB} = \left\vert \Psi\right\rangle_{A} \otimes \left\vert \Psi\right\rangle_{B} = (\sum_{i=1}\alpha_{i}\left\vert i\right\rangle_{A}) \otimes (\sum_{j=1}\beta_{j}\left\vert j\right\rangle_{B}) \text{,} \notag \\
\left\vert \Psi\right\rangle_{AB} = \left\vert \Psi\right\rangle_{A} \otimes \left\vert \Psi\right\rangle_{B} = \sum_{i,j}\ c_{ij}\left\vert i\right\rangle_{A} \left\vert j\right\rangle_{B} \text{,} \label{20}
\end{gather}%
where $c_{ij}$ are complex coeficients which satisfy a normalization condition $\sum_{i,j}\vert \ c_{ij}\vert^{2}=1$.
The state $\left\vert \Psi\right\rangle_{AB}$ is called \emph{direct product} (or separable) state if it is possible to factor it into two normalized states from the Hilbert spaces $\mathcal{H}_{A}$ and $\mathcal{H}_{B}$. A state $\left\vert \Psi\right\rangle_{AB}$ in $\mathcal{H} = \mathcal{H}_{A} \otimes \mathcal{H}_{B}$ is called \emph{entangled} if it is not a direct product state i.e., it is entangled if it cannot be factored into two normalized states elements of the two subsystems that compose the system. Entanglement describes the situation when the state of whole cannot be written in terms of 
the states of its constituent parts.
The basis for $\mathcal{H}_{A}\otimes\mathcal{H}_{B}$, where $\mathcal{H}_{A}$ and $\mathcal{H}_{B}$ are two-dimensional Hilbert spaces, is $\left\{ \left\vert 0\right\rangle_{A} \otimes \left\vert 0\right\rangle_{B}, \left\vert 0\right\rangle_{A} \otimes \left\vert 1\right\rangle_{B}, \left\vert 1\right\rangle_{A} \otimes \left\vert 0\right\rangle_{B}, \left\vert 1\right\rangle_{A} \otimes \left\vert 1\right\rangle_{B}\right\} $. 
The most general state in the Hilbert space $\mathcal{H}_{A}\otimes\mathcal{H}_{B}$ is%
\begin{gather}
\left\vert \Psi\right\rangle_{AB} = \sum_{i,j = 0}^{n=1}\ c_{ij}\left\vert i\right\rangle_{A} \left\vert j\right\rangle_{B} \text{,} \notag \\
\left\vert \Psi\right\rangle_{AB}= (c_{0}^{A} \left\vert 0\right\rangle_{A} + c_{1}^{A} \left\vert 1\right\rangle_{A}) \otimes (c_{0}^{B} \left\vert 0\right\rangle_{B} + c_{1}^{B} \left\vert 1\right\rangle_{B})\text{,} \label{21}
\end{gather}%
where $\vert c_{0}^{(A)}\vert^{2} + \vert c_{1}^{(A)}\vert^{2} = 1$ and $\vert c_{0}^{(B)}\vert^{2} + \vert c_{1}^{(B)}\vert^{2} = 1$.

\subsection{Von Neumann Entropy \\ \& Quantum Information Theory}

Entropy \cite{15,16} is the central concept of information theory. In classical physics, information processing and communication is best described by Shannon information theory. The \emph{Shannon entropy} expresses the average information we expect to gain on performing a probabilistic experiment of a random variable $A$ which takes the value $a_{i}$ with the respective probability $p_{i}$. It also can be seen as a measure of uncertainty before we learn the value of $A$. We define the Shannon entropy of a random variable $A$ by%
\begin{equation}
H(A)\equiv H(p_{1},...,p_{n})\equiv -\sum_{i=1}^{n}p_{i}\log _{2}p_{i}\text{.} \label{22}
\end{equation}%
The entropy of a random variable is completely determined by the probabilities of the different possible values that the random variable takes. Due to the fact that $p=(p_{1},...,p_{n})$ is a probability distribution, it must satisfy $\sum_{i=1}^{n}p_{i}=1$ and $0\leq p_{1},...,p_{n}\leq 1$. The Shannon entropy of the probability distribution associated with the source gives the minimal number of bits that are needed in order to store the information produced by a source, in the sense that the produced string can later be recovered. Suppose $A$ and $B$ are two random variables. The \emph{joint entropy} $H(A,B)$ measures our total uncertainty about the pair $(A,B)$. The \emph{conditional entropy} $H(A\mid B)$ is a measure of how uncertain we are about the value of $A$, given that we know the value of $B$. The \emph{mutual or correlation entropy} $H(A:B)$ measures how much information $A$ and $B$ have in common. The \emph{relative entropy} $H(p \parallel q)$ measures the closeness of two probability distributions, $p$ and $q$, defined over the same random variable $A$. The classical relative entropy of two probability distributions is related to the probability of distinguishing the two distributions after a large but finite number of independent samples (Sanov's theorem).
Since Shannon \cite{15}, \emph{information theory} or the mathematical theory of communication changed from an engineering discipline that dealed with communication channels and codes \cite{17} to a physical theory \cite{18} in where the introduction of the concepts of entropy and information were
indispensable to our understanding of the physics of measurement. Classical information theory has two primary goals \cite{19}: The first is the development of the fundamental theoretical limits on the achievable performance when communicating a given information source over a given communications channel using coding schemes from within a prescribed class. The second goal is the development of coding schemes that provide performance that is reasonably good in comparison with the optimal performance given by the theory.

The \emph{von Neumann entropy} \cite{20,21} is the quantum analogue of the Shannon's entropy. It appeared 21 years before Shannon's and generalizes Boltzmann's expression. Von Neumann  defined the entropy of a quantum state $\rho $ by the formula%
\begin{equation}
S(\rho )\equiv -Tr(\rho \ln \rho) \text{.} \label{23}
\end{equation}%
The entropy $S(\rho )$ is non-negative and takes its maximum value $\ln n$ when $%
\rho$ is maximally mixed, and its minimum value zero if $\rho$ is pure. If $\lambda _{i}$ are the eigenvalues of $\rho$ then von Neumann's definition can be expressed as $S(\rho )=-\sum_{i}\lambda _{i}\ln \lambda _{i}$. The von Neumann entropy reduces to a Shannon entropy if $\rho$ is a mixed state composed of orthogonal quantum states \cite{22}. By analogy with the Shannon entropies it is possible to define conditional, mutual and relative entropies. The negativity of the conditional entropy always indicates that two systems are entangled and indeed, how negative the conditional entropy is provides a
lower bound on how entangled the two systems are.
\emph{Quantum information theory} \cite{23,24} may be defined as the study of the achievable limits to information processing possible within quantum mechanics. Thus, the field of quantum information has two tasks: It aims to determine limits on the class of information processing tasks which are possible in quantum mechanics and provide constructive means for achieving information processing tasks. Quantum information theory appears to be the basis for a proper understanding of the emerging fields of quantum computation \cite{25,26}, quantum communication \cite{27,28}, and quantum cryptography \cite{29,30}. Entropy in quantum information theory plays prominent roles in many contexts, e.g., in studies of the classical capacity of a quantum channel \cite{31,32} and the compressibility of a quantum source \cite{33,34}.

\section{Quantum Game Theory}
For a  quantum physicist it is legitimate to ask what happens if linear superpositions of the strategies in a game are allowed  for, that is if games are generalized into the quantum domain  \cite{35}. Quantum games have demonstrated to propose a new point of view for the solution of the classical problems and dilemmas in game theory. It has been shown that quantum games are more efficient than classical games and provide a saturated upper bound for this efficiency  \cite{35,36,37,38,39,40,41,42,43,44,45}. 

In \emph{Blaquiere}'s \cite{46} \textit{Wave mechanics as a two-player game} (1980) game-theoretical ideas are discussed in the context of quantum physics. Blaquiere analyzes the connection between dynamic programming, the theory of differential games, and wave mechanics. The author argues that wave mechanics is born of a dynamic programming equation which Louis de Broglie obtained in 1923. He then expresses the stationarity principle in the form of a minimax principle written in the form of sufficiency conditions for the optimality of strategies in a two-player zero-sum differential game. The saddle-point condition, on which optimality of strategies is based, is an extension of Hamilton's principle of least action. \emph{Wiesner}'s \cite{47}  \textit{Quantum money} (1983) is believed to have started the field of quantum cryptography. Cryptographic protocols can be written in the language of game theory. Wiesner suggested to use the uncertainty principle for creating means of transmitting two messages. In 1990 \emph{Mermin} \cite{48,49} presented an n-player quantum game that can be won with certainty when it involves $n$ spin half particles in a Greenberger-Horne-Zeilinger (GHZ) state; no classical strategy can win the game with a probability greater than $1/2+1/2^{n/2}$.

The actual firsts works on quantum games were the developed for \emph{Meyer} and \emph{Eisert et al}. Meyer \cite{36} quantized a coin tossing game and found out that one player could increase his expected payoff and win with certainty by implementing a quantum strategy against his opponent's classical strategy. Eisert et al. \cite{35} developed a general protocol for two player-two strategy quantum games with entanglement by quantizing prisoner's dilemma. They found a unique Nash equilibrium, which is different from the classical one, and the dilemma could be solved if the two players are allowed to use quantum strategies. This was extended later to multiplayer games \cite{50}. \emph{Marinatto and Weber} \cite{37} extended the concept of a classical two-person static game to the quantum domain by giving a Hilbert structure to the space of classical strategies. They showed that the introduction of entangled strategies in battle of the sexes game leads to a unique solution of this game. \emph{Du et al.} \cite{39} implemented a game via nuclear magnetic resonance (NMR) system. It was demonstrated that neither of the two players would win the game if they play rationally, but if they adopt quantum strategies both of them would win. Quantum games have been used to explore unsolved problems of quantum information \cite{23,24} and in the production of algorithms for quantum computers\cite{51}. Quantum communication can be considered as a game where the objective is maximize effective communication. Also distributed computing, cryptography, watermarking and information hiding tasks can be modeled as games \cite{52,53,54,55,56,57,58}. 

\emph{Azhar Iqbal} \cite{44,45,59,60,61,62} introduced the concepts of replicator dynamics and evolutionary stable strategies from evolutionary game theory to the analysis of quantum games. \emph{Guevara}\cite{63,64,65,66,67,68,69,70} analyzed the relationships between game theory and quantum mechanics. There exists a correspondence between the replicator dynamics and the von Neumann and in general between quantum mechanics and game theory, their concepts and equilibrium definitions. \emph{Piotrowski and Sladkowski} have modeled markets, auctions and bargaining assuming traders can use quantum protocols \cite{71,72,73}. In the new quantum market games, transactions are described in terms of projective operations acting on Hilbert spaces of strategies of traders. A quantum strategy represents a superposition of trading actions that can achieve outcomes not realizable by classical means. Furthermore, quantum mechanics has features that can be used to model aspects of market behavior. For example, traders observe the actions of other players and adjust their actions and the maximal capital flow at a given price corresponds to entanglement between buyers and sellers. Nature may be playing quantum survival games at the molecular level \cite{74,75}. It could lead us to describe many of the life processes through quantum mechanics like \emph{Gogonea and Merz} \cite{76} on protein molecules. Game theory and quantum game theory offer interesting and powerful tools and their results will probably find their applications in computation, complex system analysis and cognition sciences \cite{76,77,78,79}.
 
\subsection {Quantum Strategies}

\emph{Meyer} describes his quantum penny-flip game as follows \cite{36}. The starship Enterprise faces some imminent catastrophe. Q appears on the bridge and offers P to rescue the ship if he can beat him in a penny-flip game. Q asks P to place the penny in a  small box,  head up. Then Q, P, and finally Q reaches into the box, without looking at the penny, and either flips it over or leaves it as it is. After Q's second turn they open the box  and Q wins if the penny is head up. Classically, we can take $(H,T)$ as the basis of a 2 dimensional vector space. The players moves can be represented by a
sequence of 2 $\times$2 matrices $F$ = $\begin{pmatrix}
0&1\\
1&0 \\ 
\end{pmatrix}$ (flip) and $N$ =$\begin{pmatrix}
1&0\\
0&1\\ 
\end{pmatrix}$ (not to flip) which act on a vector representing the state of the coin. A general mixed strategy is described by the matrix %
$P$ =$\begin{pmatrix}
1-p&p\\
p&1-p \\ 
\end{pmatrix}$%
, where $p \in \left[0, 1 \right]$ is the probability with which the player flips the coin. A sequence of mixed actions puts the state of the coin into a convex linear combination $aH +(1-a)T$, with $a\in \left[0, 1 \right]$. The coin is then in the state $H$ with probability $a$. Q plays his move first, after P puts the coin in the $H$ state. 
The question is what happens when Q make use of a quantum strategy, namely a sequence of unitary, rather than stochastic, matrices. The basis for $V$ is written $\left\{\left\vert H\right\rangle, \left\vert T\right\rangle\right\}$. A pure quantum state for the penny is a linear combination, where $a$, $b \in 
\mathbb{C}
$ and $\vert a\vert^{2} + \vert b\vert^{2}=1$, which means that if the box is opened, the penny will be head up with probability $\vert a\vert^{2}$. Since the penny starts in the state $\left\vert H\right\rangle$, the unitary action $U(a,b)$ by Q puts the coin into the state $a\left\vert H\right\rangle + b\left\vert T\right\rangle$, where $U(a, b)$ is $U_{1}=U(a,b)$=$\begin{pmatrix}
a&b\\
b^{*}&-a^{*}\\ 
\end{pmatrix}$%
. The initial state of the coin can be written as $\rho_{0} = \left\vert H\right\rangle \left\langle H\right\vert$. Q's action $U(a, b)$ changes the initial state $\rho_{0}$ to%
\begin{equation}
\rho_{1} = U_{1} \rho_{0}U_{1}^{\dag}=
\begin{pmatrix}
aa^{*}&ab^{*}\\
ba^{*}&bb^{*} \\ 
\end{pmatrix} \text{.}\label{24}
\end{equation}%
P's mixed action acts on this density matrix, not as a stochastic matrix on a probabilistic state, but as a convex linear combination of unitary (deterministic) transformations
\begin{gather}
\rho_{2} = pF\rho_{1}F^{\dag} + (1-p)N\rho_{1}N^{\dag}  \text{,} \notag \\
\rho_{2}=
\begin{pmatrix}
pbb^{*}+(1-p)aa^{*}&pba^{*}+(1-p)ab^{*}\\
pab^{*}+(1-p)ba^{*}&paa^{*}+(1-p)bb^{*}\\
\end{pmatrix} \text{.}\label{25}
\end{gather}%
The next move of Q, $U_{3}$ transforms the state of the penny by conjugation to $\rho_{3} = U_{3} \rho_{2}U_{3}^{\dag}$. If Q's strategy consists of $U_{1} = U(1/\sqrt{2}, 1/\sqrt{2}) = U_{3}$, his first 
action puts the penny into a simultaneous eigenvalue 1 eigenstate of both $F$ and $N$, which  is therefore invariant under any mixed strategy $pF + (1-p)N$ of P; and his second action inverts his first to give $\rho_{3} = \left\vert H\right\rangle \left\langle H\right\vert$ and wins the game. This is the optimal quantum strategy for Q.  All the pairs $([ pF+(1-p)N ]$, $[ U(1/\sqrt{2}, 1/\sqrt{2})$, $U(1/\sqrt{2}, 1/\sqrt{2})])$ are mixed/quantum equilibria for PQ Penny Flip, with value -1 to P; this is why he loses every game. The following chart explain the dynamics of the game \cite{45}%
\begin{gather}
\left\vert H\right\rangle\to \text{ }^{Q}_{\widehat{H}}\to\frac{1}{\sqrt{2}}(\left\vert H\right\rangle+\left\vert T\right\rangle)\to \text{ }^{P}_{\hat{\sigma_{x}}} \text{ or }  \text{ }^{P}_{\hat{I}} \notag \\
\to \frac{1}{\sqrt{2}}(\left\vert H\right\rangle+\left\vert T\right\rangle)\to \text{ }^{Q}_{\widehat{H}}\to \left\vert H\right\rangle \text{.} \notag 
\end{gather}%
$\widehat{H}$ is the Hadamard transformation, $\left\vert H\right\rangle$ is head, $\left\vert T\right\rangle$ is tail, $\hat{I}$ is the identity which means leaving the penny alone and $\hat{\sigma_{x}}$ flips the penny over. 
Q's quantum strategy of putting the penny into the equal superposition of head and tail, on  his first turn, means that whether $P$ flips the penny over or not, it remains in an equal superposition which Q can rotate back to head by applying the Hadamard transformation, so Q always wins when they open the box. Q's classical strategy consists of implementing $\hat{\sigma_{x}}$ or $\hat{I}$ on his turn. When Q is restricted to play only classically, flipping the penny over or not on each turn with equal probability becomes an optimal strategy for both the players. By adapting this classical strategy Q wins only with 
probability 1/2. By using a quantum strategy Q can win with probability 1.

\subsection{Classical, Evolutionary \& Quantum Prisoners Dilemma}

In this section we will analyze the famous Prisoners Dilemma from 3 points of view. It is shown how its quantization provides a solution to the problem.

\subsubsection{Prisoners Dilemma}

Two suspects, A and B, are arrested by the police but the police have insufficient evidence for a conviction. Both prisoners are separated. The police visit each of them to offer the same deal: if one testifies against the other (and the other remains silent), the betrayer goes free and the silent accomplice receives the full 10-year sentence. If both remain silent, both prisoners are sentenced to only six months in jail for a minor charge. If each betrays the other, each receives a five-year sentence. Each prisoner must take the decision of betray the other or to remain silent. But neither prisoner knows for sure what choice the other prisoner will make. The numerical values for the payoff matrix for the prisoners dilemma are chosen as in  \cite{75}, %
$ P_{A,B} = \begin{pmatrix}
(3, 3)&(0, 5)\\
(5, 0)&(1, 1) \\ 
\end{pmatrix}$. The first entry in the parenthesis denotes the payoff of A and the second is B's payoff. This choice corresponds for $r=3$ (reward), $p=1$ (punishment), $t =5$ (temptation), and $s=0$ (sucker's pay-off). 
The dilemma arises when one assumes that both prisoners only care about minimizing their own punishment. Each prisoner has only two options: to \emph{cooperate} $C$ with his accomplice and stay quiet, or to \emph{defect} $D$ from their implied pact and betray his accomplice in return for a lighter sentence. The outcome of each choice depends on the choice of the accomplice, but each prisoner must choose without knowing what his accomplice has chosen.
The catch of the dilemma is that $D$ is the dominant strategy, i.e., rational reasoning forces each player to defect, and thereby doing substantially worse than if they would both decide to cooperate. Mutual defection in prisoners dilemma is a Nash equilibrium: contemplating on the move $DD$ 
in retrospect, each of the players comes to the conclusion that he or she could not have done better by unilaterally changing his or her own strategy.

\subsubsection{Evolutionary Prisoners Dilemma}

Due to the fact that the payoff matrix for each player is the same and there are only two pure strategies, a population with two groups can be constructed. We can denote the frequency of cooperators by $x_{1}=x$, and obviously the frequency of the defectors by $x_{2}=(1-x)$, so the frequency vector would be $(\frac{x}{1-x})$ . To simplify the analysis consider the concrete example with $A =%
\begin{pmatrix}
-1&-20\\
0&-10\\ 
\end{pmatrix}$%
. This choice is analytically convenient but not exceptional, the same results will hold for any $r$, $p$, $t$, and $s$ as long as the ordering does not change. It is sufficient to study the evolution of the cooperators frequency, since the defectors frequency falls immediately from it. The average fitness of the population is  $\left\langle f(x)\right\rangle =\sum_{k,l=1}^{n=2}a_{kl}x_{k}x_{l}=a_{11}x_{1}x_{1}+a_{12}x_{1}x_{2}+a_{21}x_{2}x_{1} + a_{22}x_{2}x_{2} = 9x^{2}-10$. The fitness function for the cooperators is $f_{1} (x)=\sum_{j=1}^{n=2}a_{ij}x_{j} = a_{11}x_{1}+a_{12}x_{2}=19x-20$. The evolution for the cooperators is given by the replicator dynamics%
\begin{equation}
\dot{x}= [f_{1} - \left\langle f(x)\right\rangle] x = -(9x-10)(x-1)x \text{.}\label{26}
\end{equation}%
Note  that $ \forall x \in \left[0, 1\right]$, $x<0$. That is, the frequency of cooperators is strictly decreasing. Under the replicator dynamics the frequency of cooperators will converge to 0, leaving a population purely composed of defectors. This indicates that a population of purely defecting players is a fixed point of the system, and hence $D$ is a Nash equilibrium. The convergence to a stable point is due to the fact that the pure strategy $D$ is an ESS. Under the given dynamic, the introduction of any number of cooperators to the population will result in the extinction of those cooperators and return to the stable state. 

\subsubsection{Quantum Prisoners Dilemma}

After Meyer's work, \emph{Eisert, Wilkens and Lewenstein}  \cite{35} formulated a quantization scheme for the Prisoners Dilemma and showed that the players can escape the dilemma if they both resort to quantum strategies. Also there exists a particular pair of quantum strategies (NE) which always gives reward. There exists a particular quantum strategy which always gives at least reward if played against any classical strategy. 

The physical model consists of (i) a source of two bits, one bit for each player, (ii) a set of physical instruments which enable the player to manipulate his or her own bit in a strategic manner, and (iii) a physical measurement device which determines the players' pay-off from the state of the two bits. All three ingredients, the source, the players' physical instruments, and the pay-off physical measurement device are assumed to be perfectly known to both players. 
In this quantum formulation the classical strategies $C$ and $D$ are associated with two basis vectors $\left\vert C\right\rangle$ and $\left\vert D\right\rangle$ in the Hilbert space of two state system, i.e., a qubit. At each instance, the state of the game is described by a vector element of the tensor product space $\mathcal{H}_{A}\otimes\mathcal{H}_{B}$ which is spanned by the basis $\left\vert CC\right\rangle$, $\left\vert CD\right\rangle$, $\left\vert DC\right\rangle$, $\left\vert DD\right\rangle$, the first and second entry refers to A and B's qubit, respectively. The initial state of the game is $\left\vert \Psi_{0} \right\rangle = \hat{\mathcal{J}} \left\vert CC\right\rangle$, where $\hat{\mathcal{J}}$ is a unitary operator which is known for both players. Strategic moves of A and B are associated with unitary operators $\hat{U}_{A}$ and $\hat{U}_{B}$, respectively, which are chosen from a strategic space $S$. $\hat{U}_{A}$ and $\hat{U}_{B}$ operate exclusively on the qubits of A and B, respectively.
Having executed their moves, which leaves the game in a state $(\hat{U}_{A} \otimes \hat{U}_{B})\hat{\mathcal{J}} \left\vert CC\right\rangle$, A and B forward their qubits for the final measurement which determines their payoff. The measurement device consists of a reversible two-bit gate $\hat{\mathcal{J}}$ which is followed by a pair of Stern Gerlach type detectors. The final state of the game prior to detection is given by
\begin{equation}
\left\vert \Psi_{f}\right\rangle = \hat{\mathcal{J}}^{\dag}(\hat{U}_{A} \otimes \hat{U}_{B})\hat{\mathcal{J}} \left\vert CC\right\rangle \text{.}\label{27}
\end{equation}
The players' expected payoffs can then be written as the projections of the state $\left\vert \Psi_{f}\right\rangle$ onto the basis vectors of the tensor-product space $\mathcal{H}_{A}\otimes\mathcal{H}_{B}$. A expected payoff is given by $P_{A}=rP_{CC}+pP_{DD}+tP_{DC}+sP_{CD}, $
\begin{gather}
P_{A} = r \vert \left\langle CC\left\vert \Psi _{f}\right. \right\rangle\vert^{2} + p\vert \left\langle DD\left\vert \Psi _{f}\right. \right\rangle\vert^{2} \notag \\
+t\vert \left\langle DC\left\vert \Psi _{f}\right. \right\rangle\vert^{2}+ s\vert \left\langle CD\left\vert \Psi _{f}\right. \right\rangle\vert^{2} \text{.}\label{28}
\end{gather}
It is sufficient to restrict the strategic space to the 2-parameter set of unitary 2 $\times$ 2 matrices
\begin{equation}
\hat{U}(\theta, \phi) =%
\begin{pmatrix}
e^{i\phi}\cos\theta/2&\sin\theta/2\\
-\sin \theta/2 & e^{-i\phi}\cos\theta/2\\ 
\end{pmatrix} \label{29}
\end{equation}
with 0$\leq\theta\leq\pi$ and 0$\leq\phi\leq\pi/2$. Specificaly, $\hat{C} \equiv \hat{U}(0,0)=
\begin{pmatrix}
1&0\\
0&1\\ 
\end{pmatrix}$ and $\hat{D} \equiv \hat{U}(\pi,0)=
\begin{pmatrix}
0&1\\
-1&0\\ 
\end{pmatrix}$. $\hat{C}$ and $\hat{D}$ are the operators corresponding to the strategies of cooperation and defection respectively. To ensure that the ordinary PD is faithfully represented in its quantum version, Eisert et al. imposed additional conditions on $\hat{\mathcal{J}}$ 
\begin{equation}
\left[ \hat{\mathcal{J}}, \hat{D}\otimes\hat{D}\right] = 0\text{, } \left[\hat{\mathcal{J}}, \hat{D}\otimes\hat{C}\right]=0 \text{, } \left[\hat{\mathcal{J}}, \hat{C}\otimes\hat{D}\right]=0 \text{,}\label{30}
\end{equation}
so the operator $\hat{\mathcal{J}}$ is
\begin{equation}
\hat{\mathcal{J}}=\exp\{i\gamma\hat{D}\otimes\hat{D}/2\}
\text{,}\label{31}
\end{equation}
where $\gamma \in \left[0,\pi/2 \right]$ is a real parameter. In fact, $\gamma$ is a measure for the entanglement of the game. At $\gamma= 0$ the game reduces to its classical form. For a maximally entangled game $\gamma=\pi/2$ the classical NE, $\hat{D}\otimes\hat{D}$ is replaced by a different unique equilibrium $\hat{Q}\otimes\hat{Q}$ with $\hat{Q}\sim\hat{U}(0, \pi/2)$. The new equilibrium is also found to be Pareto optimal, that is, a player cannot increase his/her payoff by deviating from this pair of strategies without reducing the other player's payoff. Classically $(C, C)$ is Pareto optimal, but is not an equilibrium. Eisert et al. claimed that in the quantum version the dilemma in PD disappears and quantum strategies give a superior performance if entanglement is present. 

\subsection{Marinatto \& Weber's Quantum Approach}

In \emph{Quantum Approach to Static Games of Complete Information}  \cite{37} \emph{Marinatto \& Weber's} introduced a new scheme for quantizing bi-matrix games by presenting a quantum version of the Battle of Sexes. In this scheme a state in a 2 $\otimes$ 2 dimensional Hilbert space is referred to as a strategy. At the start of the game the players are supplied with this strategy. Then players manipulate the strategy. The state is finally measured and the payoffs are rewarded depending on the results of the measurement. A player can do actions within a two-dimensional subspace. Tactics are therefore local actions on a player's qubit. The final measurement, made independently on each qubit, takes into consideration the local nature of players' manipulations.

Suppose $\rho_{0}$ is the density operator of the initial strategy which the players A and B receive at the start of the game. A acts with the identity $\hat{I}$ with probability $p$ and with $\hat{\sigma_{x}}$ with probability $(1-p)$. Similarly with B. It means that each player can modify his own strategy by applying to his reduced part of the total density matrix $\rho _{0}$ %
\begin{equation}
\rho_{f}^{A(B)}= [p\hat{I}\rho_{0}^{A(B)}\hat{I}^{\dag}+(1-p)\hat{\sigma}_{x} \rho_{0}^{A(B)}\hat{\sigma}_{x}^{\dag}] \text{.}\label{32}
\end{equation}%
After the actions of the players the state changes to%
\begin{gather}
\rho_{f} = pq\hat{I}_{A}\otimes \hat{I}_{B}\rho_{0} \hat{I}_{A}^{\dag}\otimes \hat{I}_{B}^{\dag} \notag \\
+p(1-q)\hat{I}_{A} \otimes \hat{\sigma}_{xB}\rho_{0} \hat{I}^{\dag}_{A}\otimes \hat{\sigma}_{xB}^{\dag}  \notag \\
+q(1-p)\hat{\sigma}_{xA}\otimes\hat{I}_{B}\rho_{0}\hat{\sigma}_{xA}^{\dag}\otimes\hat{I}_{B}^{\dag}  \notag \\
+ (1-p)(1-q)\hat{\sigma}_{xA}\otimes \hat{\sigma}_{xB}\rho_{0}\hat{\sigma}_{xA}^{\dag}\otimes \hat{\sigma}_{xB}^{\dag} \text{.}\label{33}
\end{gather}%
In order to calculate the payoff functions $P_{A} = Tr\left\{\hat{P_{A}} \rho_{f}\right\}$ and %
$P_{B} = Tr\left\{\hat{P_{B}} \rho_{f}\right\}$ the following payoff operators were defined%
\begin{gather}
\hat{P_{A}} = \alpha_{A} \left\vert 00\right\rangle  \left\langle 00\right\vert+\beta_{A} \left\vert 01\right\rangle  \left\langle 01\right\vert + \gamma_{A} \left\vert 10\right\rangle  \left\langle 10\right\vert +\delta_{A} \left\vert 11\right\rangle \left\langle 11\right\vert \text{,} \notag
\\
\hat{P_{B}} = \alpha_{B} \left\vert 00\right\rangle  \left\langle 00\right\vert+\beta_{B} \left\vert 01\right\rangle  \left\langle 01\right\vert + \gamma_{B} \left\vert 10\right\rangle  \left\langle 10\right\vert +\delta_{B} \left\vert 11\right\rangle \left\langle 11\right\vert  \text{.}\label{34}
\end{gather}%
The scheme presented by Marinatto and Weber differs from the scheme of Eisert et al. due to the absence of the reverse gate $\hat{\mathcal{J}}$ which makes that the classical game remains as a subset of its quantum version. Also in the Marinatto and Weber scheme it is possible to get the same results one obtains in the classical version of our game. Starting from an initial state $\left\vert \Psi_{0}\right\rangle=\left\vert 00\right\rangle$ and allowing the two players to manipulate their own strategy with unitary and unimodular operators or, equivalently, through a particular transformation 
involving two hermitian operators, one interchanging states $\hat{\sigma_{x}}$ and the other leaving them unvaried $\hat{I}$. 
By the other hand, asuming that A and B have at their disposal the entangled state $\left\vert \Psi_{0}\right\rangle = a\left\vert 00\right\rangle + b\left\vert 11\right\rangle$ it is shown that both players have the same expected payoff functions making possible to choose a unique Nash equilibrium by discarding the ones which give the players the lesser reward. The entangled strategy can therefore be termed the unique solution of the quantum version of the Battle of the Sexes game.

\subsection{Quantum Evolutionary Game Theory}

Using Eisert et al. and Marinatto and Weber's quantization schemes, \emph{Iqbal and Toor} \cite{59} investigated the concept of evolutionary stable strategies within the context of quantum games and considered situations where quantization changes ESSs without affecting the corresponding Nash equilibria. 
In \emph{Evolutionary Stable Strategies in Quantum Games} they investigated the consequences when a small group of mutants using quantum strategies try to invade a classical ESS in a population engaged in a symmetric bimatrix game of prisoners dilemma. 
The classical pure strategies $C$ and $D$ are realized as $C\sim \hat{U}(0)$, $D \sim \hat{U}(\pi)$ respectively for one-parameter strategies and $C \sim \hat{U}(0, 0)$,  $D \sim \hat{U}(\pi, 0)$ respectively for two-parameter strategies. And considered three cases:  (i). A small group of mutants appear equipped with one-parameter quantum strategy $\hat{U}(\theta)$ when $D$ exists as a classical ESS. (ii). The mutants are equipped with two-parameter quantum strategy $\hat{U}(\theta, \phi)$ against classical ESS. (iii). The mutants have successfully invaded and a two-parameter quantum strategy $\hat{Q} \sim \hat{U}(0, \pi/2)$ has established itself as a new quantum ESS. Again another small group of mutants appear using some other two-parameter quantum strategy and try to invade the quantum ESS $\hat{Q}$. 
The results for these three cases were: (i). The fitness of a one-parameter quantum strategy, which also corresponds to the case of mixed (randomized) classical strategies, cannot be greater than that of a classical ESS. A one-parameter quantum strategy, therefore, cannot succeed to invade a classical ESS. 
(ii). $D$ is an ESS if $\phi < \arcsin(1/\sqrt{5})$ otherwise the strategy $\hat{U}(\theta, \phi)$ will be in 
position to invade $D$. If most of the members of the population play $D\sim \hat{U}(\pi, 0)$, then the fitness $W(D)$ will remain greater than the fitness $W [\hat{U}(\theta, \phi)]$ if $ \phi < \arcsin(1/\sqrt{5})$.  For $\phi > \arcsin(1/\sqrt{5})$ the strategy $\hat{U}(\theta, \phi)$ can invade the strategy $D$ which is an ESS. The possession of a richer strategy by the mutants in this case leads to an invasion of $D$ when $\phi > \arcsin(1/\sqrt{5})$. Mutants having access to richer strategies may seem non-judicious but even in classical setting an advantage by the mutants leading to invasion may be seen in similar context. 
(iii). A two parameter quantum strategy $\hat{U}(\theta, \phi)$ cannot invade the quantum ESS i.e., the strategy $\hat{Q} \sim \hat{U}(0,\pi/2)$ for this particular game. The mutants having access to richer strategy space remains an advantage not any more now. For the population as well as the mutants $\hat{Q}$ is the unique NE and ESS of the game. The invasion of the mutants in case (ii) does not seem so unusual given the richer structure of strategy space they exploit and they are unable to invade when it does not remain an 
advantage and most of the population have access to it. For an asymmetric quantum game between two players they showed that a strategy pair can be made an ESS for either classical (using unentangled $\left\vert \Psi_{0}\right\rangle$) or quantum (using entangled $\left\vert \Psi_{0}\right\rangle$) version of the game even when the strategy pair remains a Nash equilibrium in both the versions. They showed that in certain types of games entanglement can be used to make appear or disappear ESSs while retaining corresponding Nash equilibria. 

In \emph{Quantum mechanics gives stability to a Nash equilibrium} \cite{61}, Iqbal and Toor explored evolutionary stability in a modified Rock-Scissors-Paper quantum game. They showed that a mixed strategy NE which is not an ESS in the classical version of the game can be made an ESS when the two players play instead a quantum game by using a selected form of initial entangled state on which they apply the unitary operators in their possession. Quantum mechanics, thus, gives stability to a classical mixed NE against invasion by mutants. Stability against mutants for a mixed classical NE can be made to disappear in certain types of three player symmetric games when players decide to resort to quantum strategies. Stability against mutants in pair-wise contests coming as a result of quantum strategies have been shown a possibility for only pure strategies in certain type of symmetric games. Their results imply the selected method of quantization can bring stability against mutants to a classical mixed NE in pair-wise symmetric contests when the classically available number of pure strategies to a player is increased to three from two. A different behavior is also observed of mixed NE from pure NE in relation to quantization. 

Using again Marinatto and Weber's scheme Iqbal et al. \cite{62} analyzed the equilibria of replicator dynamics in quantum games. A well known theorem in evolutionary game theory says that an ESS is an attractor of replicator dynamics but not every attractor is an ESS. The quantization of matrix games can give or take away evolutionary stability to attractors of replicator dynamics when it is the underlying 
process of the game. They considered the effects of quantization on a saddle or a center of the dynamics. The quantization can be responsible for changing the evolutionary stability of an attractor of the dynamics.

In evolutionary game theory the Bishop-Cannings theorem does not permit pure ESSs when a 
mixed ESS exists in a bi-matrix game. However, evolutionary stability of a mixed symmetric NE 
cannot be changed with such a control. Following the approach developed for the quantum version of the rock-scissor-paper game, Iqbal et al. \cite{60} allowed the game to be played with a general form of a two-qubit initial quantum state. It becomes possible to change evolutionary stability of a mixed NE. For a bi-matrix game we worked out a symmetric mixed NE that remains intact in both the classical and quantum versions of the game. For this mixed NE they found conditions making it possible that evolutionary stability of a mixed symmetric NE changes with a switch-over of the game between its two forms, one classical and the other quantum. 

In the next sections the present chapter is concluded with the analysis of the relationships between quantum mechanics and game theory, its consequences and applications. We will propose quantization relationships for a classical game, and the quantum analogues of the replicator dynamics and the Nash equilibrium. 

\section{Relationships between Quantum Mechanics \& Game Theory}

Lets analyze some characteristic aspects of quantum mechanics and game theory.

{\scriptsize Table 1}

\begin{center}
\begin{tabular}{cc}
\hline
{\scriptsize Quantum Mechanics} & {\scriptsize Game Theory} \\ \hline
{\scriptsize n system members} & {\scriptsize n players} \\ 
{\scriptsize Quantum States} & {\scriptsize Strategies} \\ 
{\scriptsize Density Operator} & {\scriptsize Relative Frequencies Vector}
\\ 
{\scriptsize Von Neumann Equation} & {\scriptsize Replicator Dynamics} \\ 
{\scriptsize Von Neumann Entropy} & {\scriptsize Shannon Entropy} \\ 
{\scriptsize System Equilibrium} & {\scriptsize Payoff} \\ 
{\scriptsize Maximum Entropy} & {\scriptsize Maximum Payoff} \\ 
&  \\ \hline
\end{tabular}
\end{center}

Physics is a mathematical model which describes nature which usually is represented like a system composed by $n$ members. In quantum mechanics we represent and describe the state of each member through quantum states and the state of the ensemble through a density operator. The system evolves following the von Neumann equation. A measure of its order or disorder is the von Neumann entropy which is also a measure of its entanglement. The objective is the equilibrium of the system.

Game theory describes conflictive situations in populations composed by $n$ players (not necessarily people). The state of each player is given by its strategy or its relative frequencies vector which evolves in time following the replicator dynamics. The purpose of the game is a payoff and each member struggle to maximize it. 

The clear resemblances and apparent differences between both theories were a motivation to try to find an actual relationship between both systems. Due to our interests, it is important to try to analyze deeply both systems starting from its constituent elements. To start with this analysis it is important to note that the replicator dynamics is a vectorial differential equation while von Neumann equation describes the evolution of an operator (or matrix). So, if we would like to compare both systems the first we would have to do is to try to compare their evolution equations by trying to find a matrix representation of the replicator dynamics. So, the procedure we followed was the next.

\subsection{Lax Form of the Replicator Dynamics \\ \& its Properties}

As we saw the replicator dynamics is a differential equation where $x$ is a column vector. Obviously, the matrix $U=(Ax)_{i}-x^{T}Ax$ has to be diagonal and its elements are given by $u_{ii}=\sum_{j=1}^{n}a_{ij}x_{j}-\sum_{k,l=1}^{n}a_{kl}x_{k}x_{l}$. The replicator dynamics can be expressed as%
\begin{equation}
\frac{dx}{dt}=\left[ (Ax)_{i}-x^{T}Ax\right] x_{i}(t)=Ux \text{.}\label{35}
\end{equation}%
By multiplying each element of the vector $x$ by its corresponding $(x_{i})^{-1/2}$ in both parts of equation (\ref{35}) we can get%
\begin{equation}
v=U\hat{x}\text{,}\label{36}
\end{equation}%
where $v$ and $\hat{x}$ are column vectors with $v_{i}=\frac{1}{%
(x_{i})^{1/2}}\frac{dx_{i}}{dt}$ and $\hat{x}_{i}=(x_{i})^{1/2}$ respectively. Lets multiply the last equation by $\hat{x}^{T}$ and lets define the matrix%
\begin{equation}
G=\frac{1}{2}v\hat{x}^{T}=\frac{1}{2}U\hat{x}\hat{x}^{T} \text{,} \label{37}
\end{equation}%
where $g_{ij}$ are the elements of the matrix $G$, $g_{ij}=\frac{1}{2}\frac{(x_{j})^{1/2}}{(x_{i})^{1/2}}\frac{dx_{i}}{dt}=\left[\sum_{k=1}^{n}a_{ik}x_{k}-\sum_{k,l=1}^{n}a_{kl}x_{k}x_{l}\right](x_{i}x_{j})^{1/2}$. The transpose of the matrix $G$ is equal to $G^{T}=\frac{1}{2}\hat{x}v^{T}=\frac{1}{2}\hat{x}\hat{x}^{T}U^{T}$, where $g_{ij}^{T}=\frac{1}{2}\frac{(x_{i})^{1/2}}{(x_{j})^{1/2}}\frac{dx_{j}}{dt} = (x_{j}x_{i})^{1/2}\left[\sum_{k=1}^{n}a_{jk}x_{k}-\sum_{k,l=1}^{n}a_{kl}x_{k}x_{l}\right]$ are the elements of $G^{T}$.  We can define certain matrix $X$ as%
\begin{equation}
\frac{dX}{dt}=G+G^{T} =\frac{1}{2}U\hat{x}\hat{x}^{T}+\frac{1}{2}\hat{x}\hat{x}^{T}U^{T}\label{38}
\end{equation}%
so $X$ has as elements $x_{ij}=\left( x_{i}x_{j}\right) ^{1/2}$. It can be shown that the elements of $G+G^{T}$ are given by%
\begin{gather}
\left( G+G^{T}\right) _{ij} =\frac{1}{2}%
\sum_{k=1}^{n}a_{ik}x_{k}(x_{i}x_{j})^{1/2} \notag \\
+\frac{1}{2}\sum_{k=1}^{n}a_{jk}x_{k}(x_{j}x_{i})^{1/2} -\sum_{k,l=1}^{n}a_{kl}x_{k}x_{l}(x_{i}x_{j})^{1/2}\text{.}\label{39}
\end{gather}%

Lets call $\left(G_{1}\right)_{ij}=\frac{1}{2}\sum_{k=1}^{n}a_{ik}x_{k}(x_{i}x_{j})^{1/2}$, $\left(G_{2}\right)_{ij}=\frac{1}{2}\sum_{k=1}^{n}a_{jk}x_{k}(x_{j}x_{i})^{1/2}$, and $\left(G_{3}\right) _{ij}=\sum_{k,l=1}^{n}a_{kl}x_{k}x_{l}(x_{i}x_{j})^{1/2}$ the elements of the matrixes $G_{1}$, $G_{2}$ and $G_{3}$ that compose by adding the matrix $\left( G+G^{T}\right)$. 
The matrix $G_{3}$ can also be represented as $\left(G_{3}\right)_{ij}=\sum_{l=1}^{n}(x_{i}x_{l})^{1/2}\sum_{k=1}^{n}a_{kl}x_{k}(x_{l}x_{j})^{1/2}$. $G_{1}$, $G_{2}$ and $G_{3}$ can be factored in function of the matrix $X$ and the diagonal matrix $Q$, $q_{ii}=\frac{1}{2}\sum_{k=1}^{n}a_{ik}x_{k}$ so that $G_{1}=QX$, $G_{2}=XQ$ and $G_{3}=2XQX$. It is easy to show that $X^{2}=X$. We can write the equation (\ref{38}) like%
\begin{gather}
\frac{dX}{dt}=QXX+XXQ-2XQX \text{,} \notag \\
\frac{dX}{dt}=\left[ \left[ Q,X\right], X\right]=\left[ \Lambda,X\right] \text{,}\label{40}
\end{gather}%
where the matrix $\Lambda$ is $\Lambda =\left[ Q,X\right]$ and $(\Lambda )_{ij}=\frac{1}{2}\left[ \left( \sum_{k=1}^{n}a_{ik}x_{k}\right)(x_{i}x_{j})^{1/2}-(x_{j}x_{i})^{1/2}\left( \sum_{k=1}^{n}a_{jk}x_{k}\right)\right]$ are its elements. 

Matrix $X$ has the following properties: $Tr(X)=1$, $X^{2}=X$ and $X^{T}=X$. Each component of this matrix will evolve following the replicator dynamics so that we could call equation (\ref{40})  the matrix form of the replicator dynamics. 

It is easy to realize that the matrix commutative form of the replicator dynamics (\ref{40}) follows the
same dynamic than the von Neumann equation (\ref{15}). As will be shown, the properties of their correspondent elements (matrixes) are similar, being the properties corresponding to our quantum system more general than the classical system.

\subsection{Actual Relationships between Quantum Mechanics \& Game Theory }

The following table shows some specific resemblances between quantum statistical mechanics and evolutionary game theory.

{\scriptsize Table 2}

\begin{center}
\begin{tabular}{cc}
\hline
{\scriptsize Quantum Statistical Mechanics} & {\scriptsize Evolutionary Game
Theory} \\ \hline
{\scriptsize n system members} & {\scriptsize n population members} \\ 
{\scriptsize Each member in the state }$\left\vert \Psi _{k}\right\rangle $
& {\scriptsize Each member plays strategy }$s_{i}$ \\ 
$\left\vert \Psi _{k}\right\rangle $ {\scriptsize with} $p_{k}\rightarrow $
\ $\rho _{ij}${\scriptsize \ } & $s_{i}${\scriptsize \ }$\ \ \rightarrow $%
{\scriptsize \ }$\ \ x_{i}$ \\ 
$\rho ,$ $\ \ \sum_{i}\rho _{ii}{\scriptsize =1}$ & ${\scriptsize X,}$%
{\scriptsize \ \ }$\sum_{i}x_{i}{\scriptsize =1}$ \\ 
${\scriptsize i\hbar }\frac{d\rho }{dt}{\scriptsize =}\left[ \hat{H},\rho %
\right] $ & $\frac{dX}{dt}{\scriptsize =}\left[ \Lambda ,X\right] $ \\ 
${\scriptsize S=-Tr}\left\{ {\scriptsize \rho }\ln {\scriptsize \rho }%
\right\} $ & ${\scriptsize H=-}\sum_{i}{\scriptsize x}_{i}\ln 
{\scriptsize x}_{i}$ \\ 
&  \\ \hline
\end{tabular}
\end{center}

A physical system is modeled mathematically through quantum mechanics while a socioeconomical is modeled through game theory. However it is evident that both systems seem to have a similar behavior. Both are composed by $n$ members (particles, subsystems, players, states, etc.). Each member is described by a state or a strategy which has assigned a determined probability. The quantum mechanical system is described by a density operator $\rho $ whose elements represent the system average probability of being in a determined state. The socioeconomical system is described through a relative frequencies matrix $X$ whose elements represent the frequency of players playing a determined strategy. The evolution equation of the relative frequencies matrix $X$ (which describes our
socioeconomical system) is given by the Lax form of the replicator dynamics which follows the same dynamic than the evolution equation of the density operator (i.e., the von Neumann equation).
The following table shows how the properties of the matrix that describe the quantum system are more general that the properties of the matrix that describe the classical one.

{\scriptsize Table 3}

\begin{center}
$%
\begin{tabular}{cc}
\hline
{\scriptsize Density Operator} & {\scriptsize Relative freq. Matrix} \\ 
\hline
$\rho ${\scriptsize \ is Hermitian} & $X${\scriptsize \ is Hermitian} \\ 
${\scriptsize Tr\rho (t)=1}$ & ${\scriptsize TrX=1}$ \\ 
${\scriptsize \rho }^{2}{\scriptsize (t)\leq \rho (t)}$ & ${\scriptsize %
X}^{2}{\scriptsize =X}$ \\ 
${\scriptsize Tr\rho }^{2}{\scriptsize (t)\leq 1}$ & ${\scriptsize TrX}%
^{2}{\scriptsize (t)=1}$ \\ 
&  \\ \hline
\end{tabular}%
$
\end{center}

Although both systems analyzed are described through two theories apparently different both are analogous and thus exactly equivalents.
 
\section{Direct Consequences of the Relationships between Quantum Mechanics \& Game Theory}

\subsection{Quantization Relationships}

We can propose the next \textquotedblleft quantization relationships\textquotedblright %
\begin{gather}
x_{i}\rightarrow \sum_{k=1}^{n}\left\langle i\left\vert \Psi _{k}\right.
\right\rangle p_{k}\left\langle \Psi _{k}\left\vert i\right. \right\rangle
=\rho _{ii}\text{,}  \notag \\
(x_{i}x_{j})^{1/2}\rightarrow \sum_{k=1}^{n}\left\langle i\left\vert \Psi
_{k}\right. \right\rangle p_{k}\left\langle \Psi _{k}\left\vert j\right.
\right\rangle =\rho _{ij}\text{.}\label{41}
\end{gather}%
A population is represented by a quantum system in which each
subpopulation playing strategy $s_{i}$ is represented by a pure
ensemble in the state $\left\vert \Psi _{k}(t)\right\rangle$ and with
probability $p_{k}$. The probability $x_{i}$ of playing strategy $s_{i}$ or
the relative frequency of the individuals using strategy $s_{i}$ in that
population will be represented as the probability $\rho _{ii}$ of finding
each pure ensemble in the state $\left\vert i\right\rangle$  \cite{63}.

\subsection{Quantum Replicator Dynamics}

Through the last quantization relationships the replicator dynamics (in matrix commutative form) takes the form of the equation of evolution of mixed states. The von Neumann equation is the quantum analogue of the replicator dynamics. Also $X\rightarrow \rho $, $\Lambda\rightarrow -\frac{i}{\hbar }\hat{H}$, and $H(x)\rightarrow S(\rho )$  \cite{63,65}.

\subsection{Games Analysis from Quantum Information Theory}

If we define an entropy over a random variable $S^{A}$ (player's A strategic space) \cite{65} which can take the values $s_{i}^{A}$ with the respective probabilities $x_{i}^{A}$ i.e., $H(A)\equiv-\sum_{i=1}^{n}x_{i}\log _{2}x_{i}$, we could interpret the entropy of our game as a measure of uncertainty before we learn what strategy player A is going to use. If we do not know what strategy a player is going to use every strategy becomes equally probable and our uncertainty becomes maximum and it is greater while greater is the number of strategies.

If a player B decides to play strategy $s_{j}^{B}$ against player A (who plays strategy $s_{i}^{A}$) our total uncertainty about the pair $(A,B)$ can be measured by an external \textquotedblleft referee\textquotedblright\ through the joint entropy of the system $H(A,B)\equiv -\sum_{i,j}x_{ij}\log_{2}x_{ij}$, $x_{ij}$ is the joint probability to find A in state $s_{i}$ and B in state $s_{j}$. This is smaller or at least equal than the sum of the uncertainty about A and the uncertainty about B, $H(A,B)\leq H(A)+H(B)$. The interaction and the correlation between A and B reduces the uncertainty due to the sharing of information. There can be more predictability in the whole than in the sum of the parts. The uncertainty
decreases while more systems interact jointly creating a new only system.

We can measure how much information A and B share and have an idea of how their strategies or states are correlated by their mutual or correlation entropy $H(A:B)\equiv -\sum_{i,j}x_{ij}\log _{2}x_{i:j}$, with $x_{i:j}=\frac{\sum_{i}x_{ij}\sum_{j}x_{ij}}{x_{ij}}$. It can be seen easily as $H(A:B)\equiv H(A)+H(B)-H(A,B)$. The joint entropy would equal the sum of each of A's and B's entropies only in the case that there are no correlations between A's and B's states. In that case, the mutual entropy vanishes and we could not make any predictions about A just from knowing something about B.

If we know that B decides to play strategy $s_{j}^{B}$ we can determinate
the uncertainty about A through the conditional entropy $H(A\mid B)\equiv
H(A,B)-H(B)=-\sum_{i,j}x_{ij}\log _{2}x_{i\mid j}$ with $x_{i\mid j}=\frac{%
x_{ij}}{\sum_{i}x_{ij}}$. If this uncertainty is bigger or equal to zero
then the uncertainty about the whole is smaller or at least equal than the
uncertainty about A, i.e., $H(A:B)\leq H(A)$. Our uncertainty about the
decisions of player A knowing how B and C plays is smaller or at least
equal than our uncertainty about the decisions of A knowing only how B
plays $H(A\mid B,C)\leq H(A\mid B)$ i.e., conditioning reduces entropy. If
the behavior of the players of a game follows a Markov chain i.e., $%
A\rightarrow B\rightarrow C$ then $H(A)\geq H(A:B)\geq H(A:C)$ i.e., the
information can only reduces in time. Also any information C shares with A must be information which C also shares with B, $H(C:B)\geq H(C:A)$.

Two external observers of the same game can measure the difference in their
perceptions about certain strategy space of a same player A by its
relative entropy. Each of them could define a relative frequency vector, $x$
and $y$, and the relative entropy over these two probability distributions
is a measure of its closeness $H(x\parallel y)\equiv \sum_{i}x_{i}\log
_{2}x_{i}-\sum_{i}x_{i}\log _{2}y_{i}$. We could also suppose that A could
be in two possible states i.e., we know that A can play of two specific but
different ways and each way has its probability distribution (again $x$ and $%
y$ that also are known). Suppose that this situation is repeated exactly $N$
times or by $N$ people. We can made certain \textquotedblleft
measure\textquotedblright , experiment or \textquotedblleft
trick\textquotedblright\ to determine which the state of the player is. The
probability that these two states can be confused is given by the classical
or the quantum Sanov's theorem.

\subsubsection{Quantum Games Entropy}

Classically, the entropy of our system is given by $H=-Tr\left\{ X\ln X\right\}$. When the non diagonal elements of matrix $X$ are equal to zero it turns to
the Shannon entropy over the elements of the relative frequency vector $x$,
i.e., $H=-\sum_{i=1}^{n}x_{i}\ln x_{i}$. By supposing that the vector of
relative frequencies $x(t)$ evolves in time following the replicator
dynamics the evolution of the entropy of our system would be
given by%
\begin{equation}
\frac{dH}{dt}=Tr\left\{ U(\tilde{H}-X)\right\} \text{,} \label{42}
\end{equation}%
where $U_{i}=\left[ f_{i}(x)-\left\langle f(x)\right\rangle \right]$, and $%
\tilde{H}$ comes from $H=Tr\tilde{H}$.
For a quantum system the entropy is given by the von Neumann entropy which in a far from equilibrium system also vary in time until it reaches
its maximum value. When the dynamics is chaotic the variation with time of
the physical entropy goes through three successive, roughly separated stages 
\cite{25}. In the first one, $S(t)$ is dependent on the details of the
dynamical system and of the initial distribution, and no generic statement
can be made. In the second stage, $S(t)$ is a linear increasing function of
time ($\frac{dS}{dt}=const$.). In the third stage, $S(t)$ tends asymptotically towards the constant value which characterizes equilibrium ($\frac{dS}{dt}=0$). With the purpose of calculating the time evolution of
entropy we approximated the logarithm of $\rho $ by series i.e., $\ln \rho=(\rho -I)-\frac{1}{2}(\rho -I)^{2}+\frac{1}{3}(\rho -I)^{3}$... and \cite{65}%
\begin{gather}
\frac{dS(t)}{dt} =\frac{11}{6}\sum\limits_{i}\frac{d\rho _{ii}}{dt} -6\sum\limits_{i,j}\rho _{ij}\frac{d\rho _{ji}}{dt} \notag \\
+\frac{9}{2}\sum\limits_{i,j,k}\rho _{ij}\rho _{jk}\frac{d\rho _{ki}}{dt}-\frac{4}{3}\sum\limits_{i,j,k,l}\rho _{ij}\rho _{jk}\rho _{kl}\frac{d\rho _{li}}{dt}+\zeta \text{.}  \label{43}
\end{gather}%

\subsection{Thermodynamical Temperature of a Socioeconomical System}

In statistical mechanics, entropy can be regarded as a
quantitative measure of disorder. It takes its maximum possible value $\ln n$
in a completely random ensemble in which all quantum mechanical states are
equally likely and is equal to zero if $\rho $ is pure i.e., when all its
members are characterized by the same quantum mechanical state ket. 

Entropy can be maximized subject to different constraints. Generally, the result is
a probability distribution function. If we maximize $S(\rho )$ subject to the
constraints $\delta Tr\left( \rho \right) =0$ and $\delta \left\langle
E\right\rangle =0$ then%
\begin{equation}
\rho _{ii}=\frac{e^{-\beta E_{i}}}{\sum_{k}e^{-\beta E_{k}}}  \label{44}
\end{equation}%
which is the condition that the density operator must satisfy to our system
tends to maximize its entropy $S$. Without the internal energy constraint $%
\delta \left\langle E\right\rangle =0$, $\rho _{ii}=\frac{1}{N}$
which is the $\beta \rightarrow 0$\ limit (\textquotedblleft high-temperature limit\textquotedblright) in equation (\ref{44}) in where a
canonical ensemble becomes a completely random ensemble in which all energy
eigenstates are equally populated. In the opposite low-temperature limit $%
\beta \rightarrow \infty $ tell us that a canonical ensemble becomes a pure
ensemble where only the ground state is populated. The parameter $%
\beta $ is related inversely to the \textquotedblleft
temperature\textquotedblright\ $\tau $ of the system, $\beta =\frac{1}{\tau }
$. We can rewrite entropy in terms of the partition function $%
Z=\sum_{k}e^{-\beta E_{k}}$, $\beta $ and $\left\langle E\right\rangle $ via 
$S=\ln Z+\beta \left\langle E\right\rangle $. From the partition function we
can know some parameters that define the system like $\left\langle
E\right\rangle $ and $\left\langle \Delta E^{2}\right\rangle $. We can also
analyze the variation of entropy with respect to the average energy of the
system%
\begin{gather}
\frac{\partial S}{\partial \left\langle E\right\rangle }=\frac{1}{\tau }%
\text{,}  \label{45} \\
\frac{\partial ^{2}S}{\partial \left\langle E\right\rangle ^{2}}=-\frac{1}{%
\tau ^{2}}\frac{\partial \tau }{\partial \left\langle E\right\rangle }
\label{46}
\end{gather}%
and with respect to the parameter $\beta $%
\begin{gather}
\frac{\partial S}{\partial \beta }=-\beta \left\langle \Delta
E^{2}\right\rangle \text{,}  \label{47} \\
\frac{\partial ^{2}S}{\partial \beta ^{2}}=\frac{\partial \left\langle
E\right\rangle }{\partial \beta }+\beta \frac{\partial ^{2}\left\langle
E\right\rangle }{\partial \beta ^{2}}\text{.} \label{48}
\end{gather}%

\subsection{Econophysics: from Physics to Economics}

Why has it been possible to apply some methods of physics to economics and biology? It is a good reason to say that physics is a model which tries to describe phenomena and behaviors and if this model fits and describes almost exactly the observed and the measured even in the economic world then there is no problem or impediment to apply physics to solve problems in economics and biology. But, could economics, biology and physics be correlated? Could it have a relationship between
quantum mechanics and game theory? Could quantum mechanics even enclose theories like games and the evolutionary dynamics?

Based in the analysis done in this paper we can conclude that although both systems analyzed (a physical and a socioeconomical) are described through two theories apparently different (quantum mechanics and game theory) both are analogous and thus exactly equivalents. So, we could make use of some of the concepts, laws and definitions in physics for the best understanding of the behavior of economics and biology. Quantum mechanics could be a much more general theory that we had thought. From this point of view many of the equations, concepts and its properties defined quantically must be more general that its classical analogues.

It is important to note that we are dealing with very general and unspecific terms, definitions, and concepts like state, game and system. Look that the state of a system can be its strategy, and the game its behavior. Due to this, the theories that have been developed around these terms (like quantum mechanics, statistical physics, information and game theories) enjoy of this generality quality and could be applicable to model any system depending on what we want to mean for game, state, or system.
Once we have defined what system is in our model, we could try to understand what kind of 
\textquotedblleft game\textquotedblright\ is developing between its members and how they accommodate their \textquotedblleft states\textquotedblright\ in order to get their objectives and also understand the game in terms of temperature, energy, entropy, the properties and laws like if it were a physical system \cite{66,67,68,69}.

\subsection{The Collective Welfare Principle}

If our systems are analogous and thus exactly equivalents, our physical equilibrium should be also exactly equivalent to our socioeconomical equilibrium. Morerover, the natural trend of a socioeconomical system should be is to a maximum entropy state. 

Based specially on the analogous behavior between quantum mechanical systems and game theoretical systems, it is suggested the following (quantum) understanding of our (classical) system: If in an isolated system each of its accessible states do not have the same probability, the system is not in equilibrium. The system will vary and will evolve in time until it reaches the equilibrium state in where the
probability of finding the system in each of the accessible states is the same. The system will find its more probable configuration in which the number of accessible states is maximum and equally probable. The whole system will vary and rearrange its state and the states of its ensembles with the purpose of maximize its entropy and reach its equilibrium state. We could say that the purpose and maximum payoff of a physical system is its maximum entropy state. The system and its members will vary and rearrange themselves to reach the best possible state for each of them which is also the best possible state for the whole system.

This can be seen like a microscopical cooperation between quantum objects to improve their states with the purpose of reaching or maintaining the equilibrium of the system. All the members of our quantum system will play a game in which its maximum payoff is the equilibrium of the system. The members of the system act as a whole besides individuals like they obey a rule in where they prefer the welfare of the collective over the welfare of the individual. This equilibrium is represented in the maximum entropy of the system in where the system resources are fairly distributed over its members. The system is stable only if it maximizes the welfare of the collective above the welfare of the individual. If it is maximized the
welfare of the individual above the welfare of the collective the system gets unstable and eventually it collapses \emph{(Collective Welfare Principle)} \cite{63,64,65,66,67,68,69,70}.

\subsection{The Equilibrium Process called Globalization}

Lets discuss how the world process that it is called \textquotedblleft globalization\textquotedblright\ (i.e., the inexorable integration of markets, currencies, nation-states, technologies and the intensification of consciousness of the world as a whole) has a behavior exactly equivalent to a system that is tending to a maximum entropy state.

\subsubsection{Globalization: Big Communities \& Strong Currencies}

Globalization represents the inexorable integration of markets, nation-states, currencies, technologies \cite{80} and the intensification of consciousness of the world as a whole \cite{81}. This refers to an increasing global connectivity, integration and interdependence in the economic, social, technological, cultural, political, and ecological spheres \cite{82}.

Economic globalization can be measured around the four main economic flows that characterize globalization such as goods and services (e.g., exports plus imports as a proportion of national income or per capita of population), labor/people (e.g., net migration rates; inward or outward migration flows, weighted by population), capital (e.g., inward or outward direct investment as a proportion of national income or per head of population), and technology. 

Maybe the firsts of these so called big communities were the Unites States of America and the USSR (now the Russian Federation). Both consists in a set or group of different nations or states under the same basic laws or principles (constitution), objectives and an economy characterized by a same economy characterized by a same currency. Although each state or nation is a part of a big community each of them can take its own decisions and its own way of government, policies, laws and punishments (e.g., death penalty) but subject to a constitution (which is no more than a common agreement) and also subject to the decisions of a congress of the community which regulates the whole and the decisions of the parts.

The European Union stands as an example that the world should emulate by its sharing rights, responsibilities, and values, including the obligation to help the less fortunate. The most fundamental of these values is democracy, understood to entail not merely periodic elections, but also active and
meaningful participation in decision making, which requires an engaged civil society, strong freedom of information norms, and a vibrant and diversified media that are not controlled by the state or a few oligarchs. The second value is social justice. An economic and political system is to be judged
by the extent to which individuals are able to flourish and realize their potential. As individuals, they are part of an ever-widening circle of communities, and they can realize their potential only if they live in
harmony with each other. This, in turn, requires a sense of responsibility and solidarity \cite{83}.

The meeting of 16 national leaders at the second East Asia Summit (EAS) on the Philippine island of Cebu in January 2007 offered the promise of the politically fractious but economically powerful Asian mega-region one day coalescing into a single meaningful unit \cite{84}.

Seth Kaplan has offered the innovative idea of a West African Union (the 15 west african countries stretching from Senegal to Nigeria) to help solve West Africa's deep-rooted problems \cite{85}.

In South America has been proposed the creation of a Latin-American Community which is an offer for the integration, the struggle against the poverty and the social exclusion of the countries of Latin America. It is based on the creation of mechanisms to establish cooperative advantages between
countries. This would balance the asymmetries between the countries of the hemisphere and the cooperation of funds to correct the inequalities of the weak countries against the powerful nations. The economy ministers of Paraguay, Brazil, Argentina, Ecuador, Venezuela and Bolivia agreed in the
\textquotedblleft Declaraci\'{o}n de Asunci\'{o}n\textquotedblright\ to create the Bank of the South and invite the rest of countries to add to this project. The Brazilian economy minister Mantega stand out that the new bank is going to consolidate the economic, social and politic block that is appearing in South America and now they have to point to the creation of a common currency. Recently, Uruguay and Colombia have also accepted this offer and is expected the addition of more countries \cite{86}.

\subsubsection{The Equilibrium Process called Globalization}

After analyzing our systems we concluded that a socioeconomical system has a
behavior exactly equivalent that a physical system. Both systems evolve in
analogous ways and to analogous states. A system where its members are in
Nash Equilibrium (or ESS) is exactly equivalent to a system in a maximum
entropy state. The stability of the system is based on the maximization of
the welfare of the collective above the welfare of the individual. The
natural trend of a physical system is to a maximum entropy state, should not
a socioeconomical system trend be also to a maximum entropy state which
would have to be its state of equilibrium? Has a socioeconomical system
something like a \textquotedblleft natural trend\textquotedblright ?

From our analysis a population can be represented by a quantum system in
which each subpopulation playing strategy $s_{i}$ is represented by a
pure ensemble in the state $\left\vert \Psi _{k}(t)\right\rangle $ and with
probability $p_{k}$. The probability $x_{i}$ of playing strategy $s_{i}$ or
the relative frequency of the individuals using strategy $s_{i}$ in that
population is represented as the probability $\rho _{ii}$ of finding
each pure ensemble in the state $\left\vert i\right\rangle $. Through these
quantization relationships the replicator dynamics takes the form of the equation of evolution of mixed states. The von Neumann equation is the quantum analogue of  the replicator dynamics.

Our now \textquotedblleft quantum statistical system\textquotedblright\ 
composed by quantum objects represented by quantum states which represent
the strategies with which \textquotedblleft players\textquotedblright\
interact is characterized by certain interesting physical parameters like
temperature, entropy and energy.

In this statistical mixture of ensembles (each ensemble characterized by
a state and each state with a determined probability) its natural
trend is to its maximum entropy state. If each of its accessible states do
not have the same probability, the system will vary and will evolve in time
until it reaches the equilibrium state in where the probability of finding
the system in each of the accessible states is the same and its number is
maximum. In this equilibrium state or maximum entropy state the system
\textquotedblleft resources\textquotedblright\ are fairly distributed over
its members. Each ensemble is equally probable, is characterized
by a same temperature and by a stable state.

Socioeconomically and based on our analysis, our world could be understood
as a statistical mixture of \textquotedblleft ensembles\textquotedblright\
(countries for example). Each of these ensembles are characterized by a
determined state and a determined probability. But more important, each
\textquotedblleft country\textquotedblright\ is characterized by a specific
\textquotedblleft temperature\textquotedblright\ which is a measure of the socioeconomical activity of that ensemble. That temperature is
related with the interactions between the members of
the ensemble. The system will evolve naturally to a maximum entropy state.
Each pure ensemble of this statistical mixture will vary and accommodate its
state until get the \textquotedblleft thermal equilibrium\textquotedblright\ .
First with its nearest neighbors creating new big ensembles characterized
each of them by a same temperature. Then through the time, these new big
ensembles will seek its \textquotedblleft thermal
equilibrium\textquotedblright\ between themselves and with its nearest
neighbors\ creating new bigger ensembles. The system will continue evolving
naturally in time until the whole system get an only state characterized by
a same \textquotedblleft temperature\textquotedblright.

This behavior is very similar to what has been called globalization. The
process of equilibrium that is absolutely equivalent to a system that is
tending to a maximum entropy state is the actual globalization. This
analysis predicts the apparition of big common \textquotedblleft
markets\textquotedblright\ or (economical, political, social, etc.)
communities of countries (European Union, Asian Union, Latin-American
Community, African Union, Mideast Community, Russia and USA) and strong
common currencies (dollar, euro, yen, sol, etc.). The little and poor
economies eventually will be unavoidably absorbed by these \textquotedblleft
markets\textquotedblright\ and these currencies. If this process continues
these markets or communities will find its \textquotedblleft
equilibrium\textquotedblright\ by decreasing its number until reach a state
in where there exists only one big common community (or market) and only one
common currency around the world \cite{70}.

\section{Conclusion}

We have reviewed the evolution of games analysis from classical through evolutionary and quantum game theory. From the basic concepts and definitions for 2 players we have extended them to a population and its dynamics. We analyzed the equilibria, Nash equilibrium, the evolutionary stable strategies and the collective welfare principle.  And some new concepts in game theory like entanglement, entropy and temperature.

Moreover, the relationships between quantum mechanics and game theory and its direct consequences and applications. The correspondence between the replicator dynamics  and the von Neumann equation and between the NE and the Collective Welfare Principle. And how quantum mechanics (and physics) could be used to explain more correctly biological and economical processes (econophysics). 

Finally, we presented an interesting result consequence from our analysis which proves that the so called  \textquotedblleft globalization\textquotedblright\ process (i.e., the inexorable integration of markets, currencies, nation-states, technologies and the intensification of consciousness of the world as a whole) has a behavior exactly equivalent to a system that is tending to a maximum entropy state. This let us predict the apparition of big common markets and strong common currencies that will reach the  \textquotedblleft equilibrium\textquotedblright\ by decreasing its number until they get a state characterized by only one common currency and only one big common community around the world.

\begin{acknowledgments}
The author (Esteban Guevara) undertook this work with the support of the
Abdus Salam International Centre for Theoretical Physics ICTP at Trieste,
Italy and its Programme for Training and Research in Italian Laboratories
ICTP-TRIL which let him continue with his research at the Center for
Nonlinear and Complex Systems at Universit\`{a} degli Studi dell'Insubria in
Como.
\end{acknowledgments}

\end{document}